\documentclass[12pt,aps,preprint,nofootinbib]{revtex4}

\usepackage{graphicx}
\usepackage{cancel}
\usepackage{amssymb}
\usepackage{textcomp}
\usepackage{amsmath}
\usepackage{bm}
\usepackage{times}
\usepackage{epsfig}
\usepackage{epstopdf}
\usepackage{color}
\usepackage{multirow}
\usepackage[colorlinks=true, pdfstartview=FitV, linkcolor=blue, citecolor=blue, urlcolor=blue]{hyperref}
\usepackage{slashbox}

\DeclareMathAlphabet{\mathcal}{OMS}{zplm}{m}{n}
\DeclareMathAlphabet{\pazocal}{OML}{zplm}{m}{n}

\def\lsim{\mathrel{\raise.3ex\hbox{$<$\kern-.75em\lower1ex\hbox{$\sim$}}}}
\def\gsim{\mathrel{\raise.3ex\hbox{$>$\kern-.75em\lower1ex\hbox{$\sim$}}}}

\def\beq{\begin{equation}}
\def\eeq{\end{equation}}
\def\be{\begin{equation}}
\def\ee{\end{equation}}
\def\bea{\begin{eqnarray}}
\def\eea{\end{eqnarray}}

\begin{document}



\title{Dark matter properties implied by gamma ray interstellar emission models}

\author{Csaba Bal\'azs$^{1,2,}$} \email{csaba.balazs@monash.edu}
\author{Tong Li$^{1,}$} \email{tong.li@monash.edu}

\affiliation{$^1$
ARC Centre of Excellence for Particle Physics at the Tera-scale \\
School of Physics and Astronomy, Monash University \\
Melbourne, Victoria 3800, Australia}
\affiliation{$^2$
Monash Centre for Astrophysics, Monash University, Melbourne, Victoria 3800, Australia}

\begin{abstract}

We infer dark matter properties from gamma ray residuals extracted using eight different interstellar emission scenarios proposed by the Fermi-LAT Collaboration to explain the Galactic Center gamma ray excess.  Adopting the most plausible simplified ansatz, we assume that the dark matter particle is a Majorana fermion interacting with standard fermions via a scalar mediator.  Using this theoretical hypothesis and the Fermi residuals we calculate Bayesian evidences, including Fermi-LAT exclusion limits from 15 dwarf spheroidal galaxies as well.  Our Bayes factors single out four of the Fermi scenarios as compatible with the simplified dark matter model.  In the most preferred scenario the dark matter (mediator) mass is in the 100-500 (1-200) GeV range and its annihilation is dominated by top quark final state.  Less preferred but still plausible is annihilation into $b\bar{b}$ and $\tau^+\tau^-$ final states with an order of magnitude lower dark matter mass.  Our conclusion is that the properties of dark matter extracted from gamma ray data are highly sensitive to the modeling of the interstellar emission.
\end{abstract}


\maketitle

\section{Introduction}

Based on the data collected by the Large Area Telescope (LAT) on board the Fermi Gamma Ray Space Telescope an excess of gamma rays has been found~\cite{Goodenough:2009gk, Hooper:2010mq, Boyarsky:2010dr, Abazajian:2012pn, Hooper:2013rwa, Gordon:2013vta, Daylan:2014rsa, Calore:2014xka}.  The excess photons, appearing in the energy spectrum of the gamma ray flux around 2 GeV, originate from an extended volume centered on the Galactic Center (GC) and their source is still under scrutiny \cite{Fornasa:2016ohl, Gomez-Vargas:2013bea, Abazajian:2014fta, Ipek:2014gua, Ko:2014gha, Cline:2014dwa, Ko:2014loa, arXiv:1405.0272, Kong:2014haa, Boehm:2014bia, Ghosh:2014pwa, Wang:2014elb, Fields:2014pia, Berlin:2014tja, Carlson:2014cwa, Petrovic:2014uda, arXiv:1312.7488, arXiv:1402.4500, arXiv:1402.6671, arXiv:1403.1987, arXiv:1403.3401, arXiv:1403.5027, arXiv:1404.1373, arXiv:1404.2018, arXiv:1404.2067, arXiv:1404.2318, arXiv:1404.2572, arXiv:1404.3362, arXiv:1404.5503, arXiv:1404.6528, arXiv:1405.1030, arXiv:1405.1031, arXiv:1405.5204, arXiv:1405.4877, arXiv:1405.6240, arXiv:1405.6709, arXiv:1405.7059, arXiv:1405.7370, arXiv:1406.0507, arXiv:1406.1181, arXiv:1406.2276, arXiv:1406.4683, arXiv:1406.5662, arXiv:1406.6372, arXiv:1406.6408, arXiv:1408.4929, arXiv:1408.5795, arXiv:1409.1406, arXiv:1409.7864, arXiv:1410.3239, arXiv:1410.4842, arXiv:1411.2592, arXiv:1411.2619, arXiv:1411.4647, arXiv:1412.1663, arXiv:1412.5174, arXiv:1501.00206, arXiv:1501.07275, arXiv:1501.07413, arXiv:1502.05682, arXiv:1502.05703, arXiv:1503.06348, arXiv:1503.08213, arXiv:1503.08220, arXiv:1504.03610, arXiv:1504.03908, arXiv:1505.04620, arXiv:1505.04988, arXiv:1505.07826, arXiv:1506.05119, arXiv:1507.01793, arXiv:1507.02288, arXiv:1507.04644, arXiv:1507.07008, arXiv:1507.07922, arXiv:1507.08295, arXiv:1507.05616, arXiv:1508.05716, Dutta:2015ysa, arXiv:1509.02928, arXiv:1509.05076, arXiv:1509.09050, arXiv:1510.00714, arXiv:1510.04698, arXiv:1510.06424, arXiv:1510.07562, arXiv:1511.02938, arXiv:1511.09247, arXiv:1512.00698, arXiv:1512.01846, arXiv:1512.02899, arXiv:1512.04966, arXiv:1512.06825, arXiv:1601.05089, arXiv:1601.05797, arXiv:1602.00590, arXiv:1602.04788, arXiv:1602.05192, arXiv:1603.08228, arXiv:1604.00744, arXiv:1604.01039, arXiv:1604.01026, arXiv:1604.01402, arXiv:1604.04589, arXiv:1604.06566, arXiv:1606.09250, arXiv:1607.00737, arXiv:1608.00786, arXiv:1608.07289}.
The primary Galactic diffuse emission components of gamma rays arise from the interaction of Galactic cosmic rays with interstellar gas and radiation fields in the Milky Way.
Additional gamma ray components are believed to arise from individual point sources, isotropic gamma ray background and possibly extra Galactic Center sources such as dark matter (DM).

The Fermi-LAT Collaboration recently released their analysis of data taken during the first 62 months of observation in the direction of the Galactic Center~\cite{TheFermi-LAT:2015kwa}.
In their analysis four customized interstellar emission models (IEMs) and various point sources in different catalogs~\cite{Abdo:2010ru,Fermi-LAT:2011yjw,Acero:2015hja} were included.  After subtracting the
diffuse emission and point source contributions from the data they found residuals which are compatible with spatial templates produced by dark matter particle annihilation, characterized by two different choices of spectral models, in the Galaxy.  As a crucial importance, the Fermi fit indicated that the potential dark matter component strongly depends on IEM assumptions.  This dependence leads to an uncertainty that propagates into dark matter properties extracted from the gamma ray data.

To quantify the dependence of inferred dark matter properties on the IEM assumptions, following the Fermi-LAT Collaboration, we assume that dark matter annihilation is the source of the Galactic Center gamma ray excess.  To endow dark matter particles with specific properties we use a straw man dark matter model.  To make minimal and general theoretical assumptions we use the simplified model framework assuming that dark matter is a Majorana fermion coupling to standard fermions via a scalar mediator.  This particular simplified model has unsuppressed indirect, and suppressed direct and collider detection rates, and emerged as the most plausible one when compared with a wide range of data \cite{Balazs:2014jla, Balazs:2015boa, Balazs:2015iwa}.  To show the role of the IEM assumptions in constraining dark matter properties, we infer the properties of these dark matter particles using the multiple Fermi data sets categorized by different choices of IEMs and spectral models.

To remain pragmatic, additionally to the Galactic Center data, we impose constraints on the dark matter model coming from dwarf spheroidal satellite galaxies (dSphs) of the Milky Way.  These dwarf galaxies are known to be dominated by dark matter.  The Fermi-LAT Collaboration recently set upper limits on the dark matter annihilation cross section from a combined analysis of 15 Milky Way dSphs~\cite{Ackermann:2015zua}.  These constraints are the most stringent for dark matter annihilating into quark or $\tau$ lepton channels, compared to constraints from cosmic microwave background (CMB) and AMS-02~\cite{Elor:2015bho}.  Thus, we include the Fermi dSphs limit in our analysis.

The rest of our paper is organized as follows.  In Sec.~\ref{sec:Models}, we outline the simplified dark matter model we consider.  In Sec.~\ref{sec:Constraints}, we define the observables, gamma ray flux from the Galactic Center and dwarf galaxies, which we include in a composite likelihood function.  Our numerical results are given in Sec.~\ref{sec:Results}, and we present our conclusions in Sec.~\ref{sec:Concl}.  The statistical background of our analysis is summarized in the Appendix.

\section{The dark matter model}
\label{sec:Models}

In this section, we describe the simplified dark matter model we use in our analysis.
Majorana fermions recently emerged as the most plausible dark matter particle candidates in the simplified model context \cite{Balazs:2014jla, arXiv:1407.1859, arXiv:1408.2223, arXiv:1409.5776, arXiv:1501.03164, arXiv:1503.01500, Balazs:2015boa, arXiv:1507.02288, Balazs:2015iwa}.  Motivated by this, we assume that dark matter consists of a single Majorana fermion, which we denote by $\chi$.  Additionally, motivated by the listed literature and the Higgs portal, we assume that the interaction between the dark matter particle and standard fermions is mediated by a scalar particle $S$ \cite{arXiv:1306.4710, deS.Pires:2010fu}.

Following Refs. \cite{Balazs:2014jla, Balazs:2015boa, Balazs:2015iwa}, we describe the dark matter to mediator interaction by
\begin{eqnarray}
{\cal L}_\chi \supset \frac{i\lambda_\chi}{2}\bar{\chi}\gamma_5\chi S ,
\label{eq:interaction0}
\end{eqnarray}
and set $\lambda_\chi=1$.
This is a typical choice to reduce the dimensions of the theoretical parameter space by effectively absorbing $\lambda_\chi$ into the couplings of the mediator to standard model fermions $f$.  The latter enters the Lagrangian via
\begin{eqnarray}
{\cal L}_S \supset \lambda_f \bar{f}f S.
\label{eq:intercation1}
\end{eqnarray}
We only assume couplings between $S$ and third generation fermions $f=b,t,\tau$ to remain consistent with minimal flavor violation~\cite{D'Ambrosio:2002ex}.
%

Our main motivation for the form of the interactions defined in Eqs.~(\ref{eq:interaction0}) and (\ref{eq:intercation1}) is to avoid velocity suppressed dark matter annihilation~\cite{Kumar:2013iva}.
The $\gamma_5$ in Eq.~(\ref{eq:interaction0}) lifts velocity suppression thus boosting the plausibility of  this model to explain the gamma ray excess.  Additionally, for this scenario, spin dependent dark matter-nucleon elastic scattering cross section is forbidden and the spin independent cross section is momentum suppressed ensuring immunity to direct detection limits.
In summary, the following masses and couplings span the dark matter particle model parameter space:
\begin{eqnarray}
{\pazocal p} = \left\{ m_\chi, m_S, \lambda_b, \lambda_t, \lambda_\tau \right\} .
\label{param}
\end{eqnarray}

\section{Observables}
\label{sec:Constraints}

\subsection{Gamma ray flux from the Galactic Center}

When modeling the distribution of gamma rays from the Galactic Center in Ref. \cite{TheFermi-LAT:2015kwa} the Fermi-LAT collaboration includes a contribution produced by annihilation (or decay) of dark matter particles.  In our scenario this contribution comes from the self-annihilating $\chi$ particles.  The energy distribution of this gamma ray emission component is
\begin{eqnarray}
\frac{d\Phi_\gamma}{dE}=\frac{\langle \sigma v\rangle}{8\pi m_\chi^2}J\sum_{f=b,t,\tau} B_f \frac{dN_\gamma^f}{dE}.
\label{flux}
\end{eqnarray}
The first term on the right hand side carries the dependence on the dark matter model via the (velocity averaged) annihilation cross section $\langle \sigma v\rangle$ of the dark matter particles near the Galactic Center, together with the dark matter mass $m_\chi$ and $B_f$, the annihilation fraction $\langle \sigma v\rangle_f/ \langle \sigma v\rangle$ into the $f{\bar f}$ final state.  The energy distribution of photons $dN_\gamma^f/dE$ produced in the annihilation channel with final state $f{\bar f}$ also comes from particle physics, although it is dark matter model independent.  
The $J$ factor is defined through an integral of the Galactic dark matter distribution
\begin{eqnarray}
J=4\int\rho_\chi^2(r)dx\cos(b)dbd\ell .
\end{eqnarray}
In terms of the Galactic latitude $b$ and longitude $\ell$ the radial variable is
\begin{eqnarray}
r=\sqrt{x^2+r_\odot^2-2xr_\odot \cos(b)\cos(\ell)} .
\end{eqnarray}

To match Ref. \cite{TheFermi-LAT:2015kwa} we use the generalized Navarro-Frenk-White (NFW) dark matter profile \cite{astro-ph/9508025} describing the dark matter spatial distribution in the Galaxy
\begin{eqnarray}
\rho_\chi(r)=\rho_0\frac{(r/r_s)^{-\gamma}}{(1+r/r_s)^{3-\gamma}} ,
\end{eqnarray}
with $r_s=20$ kpc, $r_\odot=8.5$ kpc, and $\rho_\chi(r_\odot)=0.3 \ {\rm GeV/cm^3}$.  Following the earlier literature using preliminary Fermi results, we fix the inner slope of the NFW halo profile to $\gamma=1.2$~\cite{Agrawal:2014oha,Cline:2015qha}.  When calculating the $J$ factor we integrate over a $15^\circ\times 15^\circ$ region centered on the GC.


The total differential yield in Eq.~(\ref{flux}) is the sum of partial differential yields $dN_\gamma^f/dE$ weighted by the annihilation fractions $B_f$.  Each partial differential yield is coming from pairs of dark matter particles annihilating into a specific fermionic final state $f$ which, via fragmentation or hadronization, produces photons.  The sum over the differential yields runs over three third generation, charged SM fermions: $b,t,\tau$.  The shapes of the three differential yields we consider are very different as shown by FIG. 1 of Ref.~{\cite{Balazs:2014jla}}.  Since $B_f$ depends on the model parameters listed in Eq.~(\ref{param}), the gamma ray data is very important to constraining these, especially the coupling of the mediator particle to SM fermions.  We used micrOmegas (version 3.6.9) to calculate the full differential gamma ray flux~\cite{arXiv:1305.0237}.

To account for the primary source of gamma rays from the Galactic Center the Fermi-LAT Collaboration considered different scenarios for the Galactic diffuse emission.  They used the public code GALPROP to calculate the IEM component corresponding to each of these cases.  Modeling the spatial distribution of cosmic ray sources in GALPROP is based on the observed supernova remnants (SNR)~\cite{Case:1998qg}, pulsars~\cite{Yusifov:2004fr} or OB (O or early-type B spectral) type stars~\cite{Bronfman:2000tw}.  These prefer a non-vanishing pulsar distribution near the Galactic Center and OB stars distributed mostly on the Galactic disk~\cite{Carlson:2016iis}.  The Fermi-LAT Collaboration thus selected their IEMs based on extreme distributions, i.e. intensity-scaled pulsars, intensity-scaled OB-stars, index-scaled pulsars and index-scaled OB-stars.  Point source candidates were then combined with the above IEMs in the Fermi-LAT analysis.  In order to account for the observed residuals, the Fermi-LAT Collaboration included the NFW profile as the spatial template to model the additional gamma ray distribution from dark matter annihilation.  They used two NFW spectral models: power law per energy band and exponential cut-off power law \cite{TheFermi-LAT:2015kwa}.  The eight resulting IEM scenarios are listed in Table \ref{tab:FermiScenarios}.

\begin{table}[h]
\begin{center}
\setlength\tabcolsep{2mm}
\begin{tabular}{r|l}
Fermi scenario & assumptions \\
\hline
IEM1 & intensity-scaled pulsars, exponential cut-off NFW \\
IEM2 & intensity-scaled OB stars, exponential cut-off NFW \\
IEM3 & index-scaled pulsars, exponential cut-off NFW \\
IEM4 & index-scaled OB stars, exponential cut-off NFW \\
IEM5 & intensity-scaled pulsars, per E-band NFW \\
IEM6 & intensity-scaled OB stars, per E-band NFW \\
IEM7 & index-scaled pulsars, per E-band NFW \\
IEM8 & index-scaled OB stars, per E-band NFW \\
\end{tabular}
\end{center}
\caption{Fermi scenarios for interstellar emission models (IEMs) that we adopt.  The corresponding Galactic gamma ray excess is shown in Figures 12, 13, 17 and 18 of Ref.~\cite{TheFermi-LAT:2015kwa}.}
\label{tab:FermiScenarios}
\end{table}

Using the eight IEM scenarios listed in TABLE~\ref{tab:FermiScenarios}, the Fermi-LAT Collaboration extracted differential gamma ray fluxes quantifying the excess from the Galactic Center.  They presented corresponding upper and lower limits, reflecting their fit uncertainties, in Figure 13 and 18 of Ref.~\cite{TheFermi-LAT:2015kwa}.
As experimental data for the Galactic Center gamma ray excess, we take these eight distributions.  We assume that the ($1 \sigma$) uncertainty of each distribution is given by the width of the band presented in Ref.~\cite{TheFermi-LAT:2015kwa}.  Divided into 20 energy bins we include the output of the Fermi fit, together with predictions of the above dark matter model, in a composite likelihood function.
The general form of our likelihood function, for the eight analyzed gamma ray source models, is given by
\begin{eqnarray}
\mathcal{L}_{{\rm IEM}i} = \prod_{j_{\rm bin}=1}^{20}\mathcal{L}^{\rm Gauss} \left(\left.\left.\frac{d\Phi_\gamma}{dE}\right|_{\mathrel{^{{\rm IEM}i}_{j_{\rm bin}}}}\right|\pazocal{p}\right), \ \ \ {\rm IEM}i={\rm IEM}1,...,{\rm IEM}8 .
\label{eq:LGalaCent}
\end{eqnarray}
Here $\mathcal{L}^{\rm Gauss}$ has a Gaussian form which is given in the Appendix, and the product runs over 20 energy bins for each IEM$i$ scenario.  The likelihood function depends on the extracted gamma ray flux residual given by Eq.~(\ref{flux}), and (via the theoretical prediction for the gamma ray flux) the parameter set $\pazocal{p}$ given in Eq.~(\ref{param}).

\subsection{Gamma ray flux from Dwarf Galaxies}

Signs of dark matter annihilation were searched for in gamma rays originating from Milky Way satellite dwarf galaxies in six years of Fermi-LAT data, but no significant signal was detected~\cite{Ackermann:2015zua}.  The gamma ray flux for this case is given by
\begin{eqnarray}
E\Phi_\gamma=\frac{\langle \sigma v\rangle}{8\pi m_\chi^2}J \sum_{f=b,t,\tau} B_f \int_{E_{\rm min}}^{E_{\rm max}} \frac{dN_\gamma^f}{dE}EdE.
\label{eq:dwarfflux}
\end{eqnarray}
The Fermi-LAT Collaboration analyzed 25 Milky Way dwarf galaxies and split the observed energy region ($0.5-500$ GeV) into 24 bins for the integral in Eq.~(\ref{eq:dwarfflux}) for each galaxy.  They included 15 of these dwarfs in their combined analysis (the ones above the middle horizontal line in TABLE I of \cite{Ackermann:2015zua}).  In each energy bin they released an upper limit on $E\Phi_\gamma$ for all analyzed dwarf galaxies.  We include these upper limits in our likelihood function and take 100\% of the theoretical prediction as the uncertainty following the expected sensitivity in Ref.~\cite{Ackermann:2015zua}.  For the $J$ factors of dwarf galaxies, we adopt the observed ones listed in Table I of Ref.~\cite{Ackermann:2015zua} only keeping the 15 which were included in their combined analysis.

Our likelihood function for the dwarf galaxies is given by
\begin{eqnarray}
\mathcal{L} = \prod_{\rm dwf=1}^{15}\prod_{j_{\rm bin=1}}^{24} \mathcal{L}_{\rm dwf}^{\rm Err} \left.\left(\left. E \Phi_\gamma (J_{\rm dwf}) \right|_{j_{\rm bin}} \right| {\pazocal p}\right) ,
\label{eq:Ldwarf}
\end{eqnarray}
with $\mathcal{L}_{\rm dwf}^{\rm Err}$ being an error function and defined in detail in the Appendix.  Here we include the 15 dwarf galaxies with 24 energy bins for each of them.  The likelihood function depends on the upper limits set by the gamma ray flux, the corresponding $J$ factors, $J_{\rm dwf}$, and the dark matter model parameter set $\pazocal{p}$.

\section{Results}
\label{sec:Results}

We implemented the simplified dark matter model, described in Sec.~\ref{sec:Models}, in FeynRules~\cite{arXiv:1310.1921} and generated input model files for micrOmegas~\cite{arXiv:1305.0237}.  The calculation of the gamma ray fluxes was performed by a custom version of micrOmegas (based on version 3.6.9).  We weighted the composite likelihood function, the product of Eqs.~(\ref{eq:LGalaCent}) and (\ref{eq:Ldwarf}), with priors given in TABLE~\ref{tab:Priors}.
\begin{table}[ht]
\setlength\tabcolsep{2mm}
\begin{center}
\begin{tabular}{|c|c|c|c|c|c|}
\hline
DM model parameter & $m_\chi$ (GeV) & $m_S$ (GeV) &  $\lambda_b$ & $\lambda_t$ & $\lambda_\tau$ \\
\hline
scan range & $1-10^3$ & $1-10^3$ & $10^{-5}-10$ & $10^{-5}-10$ & $10^{-5}-10$ \\
\hline
prior type & log & log & log & log & log \\
\hline
\end{tabular}
\end{center}
\caption{Priors of the theory parameters we scanned.}
\label{tab:Priors}
\end{table}
Afterwards, we sampled the posterior distribution over the theoretical parameter space using MultiNest~\cite{arXiv:0809.3437} using 5000 live points and requiring a tolerance factor of 0.5.  The estimated integral of the posterior distributions, the model evidence, is an output of MutliNest.  Using this output, we constructed evidence ratios, Bayes factors
\begin{eqnarray}
{\cal B} = \frac{{\cal E}({\rm IEM}i)}{{\cal E}({\rm IEM}j)} ,
\label{eq:BayesFactor}
\end{eqnarray}
for every combination of the eight Fermi IEM scenarios.  We give the definition of the evidence $\cal E$ in the Appendix.  These Bayes factors, the main result of our analysis, are presented in Table~\ref{bayes}.





\begin{table}[ht]
\setlength\tabcolsep{2mm}
\begin{tabular}{|c|c|c|c|c|c|c|c|c|}
\hline
\backslashbox{~~~IEM$i$}{IEM$j$~~~} & IEM1 & IEM2 & IEM3 & IEM4 & IEM5 & IEM6 & IEM7 & IEM8 \\ \hline
IEM1 & $0$ & $-1.5$ & $+2.4$ & $+15$ & $+57$ &  $+52$ & $+6.9$ & $+34$ \\ \hline
IEM2 &     &    $0$ & $+3.9$ & $+16$ & $+59$ &  $+53$ & $+8.4$ & $+36$ \\ \hline
IEM3 &     &        &    $0$ & $+12$ & $+55$ &  $+50$ & $+4.5$ & $+33$ \\ \hline
IEM4 &     &        &        &   $0$ & $+43$ &  $+37$ & $-7.6$ & $+20$ \\ \hline
IEM5 &     &        &        &       &   $0$ & $-5.5$ &  $-51$ & $-23$ \\ \hline
IEM6 &     &        &        &       &       &    $0$ &  $-45$ & $-18$ \\ \hline
IEM7 &     &        &        &       &       &        &    $0$ & $+27$ \\ \hline
IEM8 &     &        &        &       &       &        &        &   $0$ \\ \hline
\end{tabular}
\caption{The natural logarithm of model evidence ratios, Bayes factors, for the eight Fermi IEM scenarios calculated using the simplified dark matter model presented in Sec.~\ref{sec:Models}, including Fermi-LAT limits from 15 dwarf Milky Way satellite galaxies.  (By definition the table is antisymmetric so we omitted to show elements below the diagonal.)}
\label{bayes}
\end{table}

\begin{figure}[t]
\begin{center}
\includegraphics[scale=1,width=5cm]{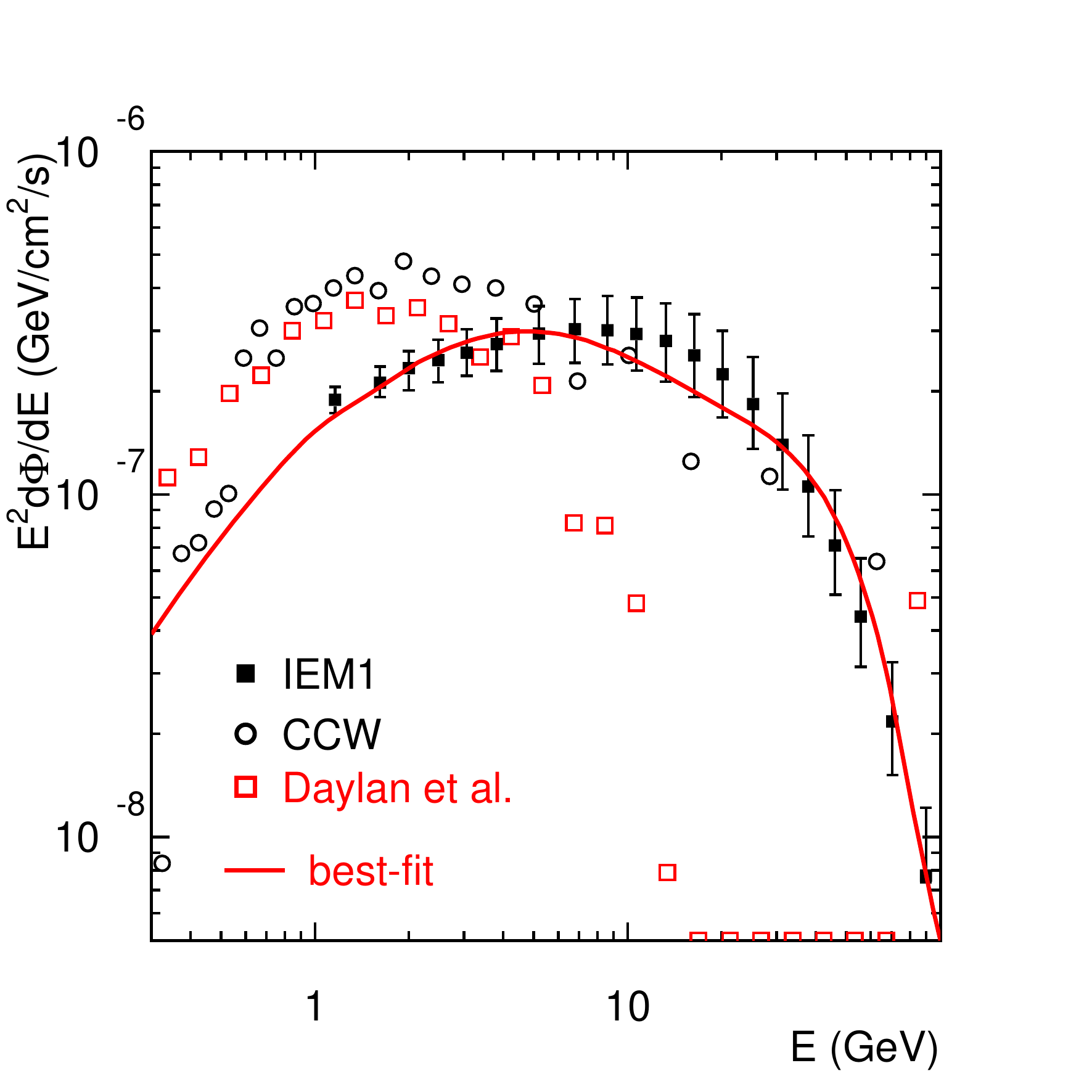}
\includegraphics[scale=1,width=5cm]{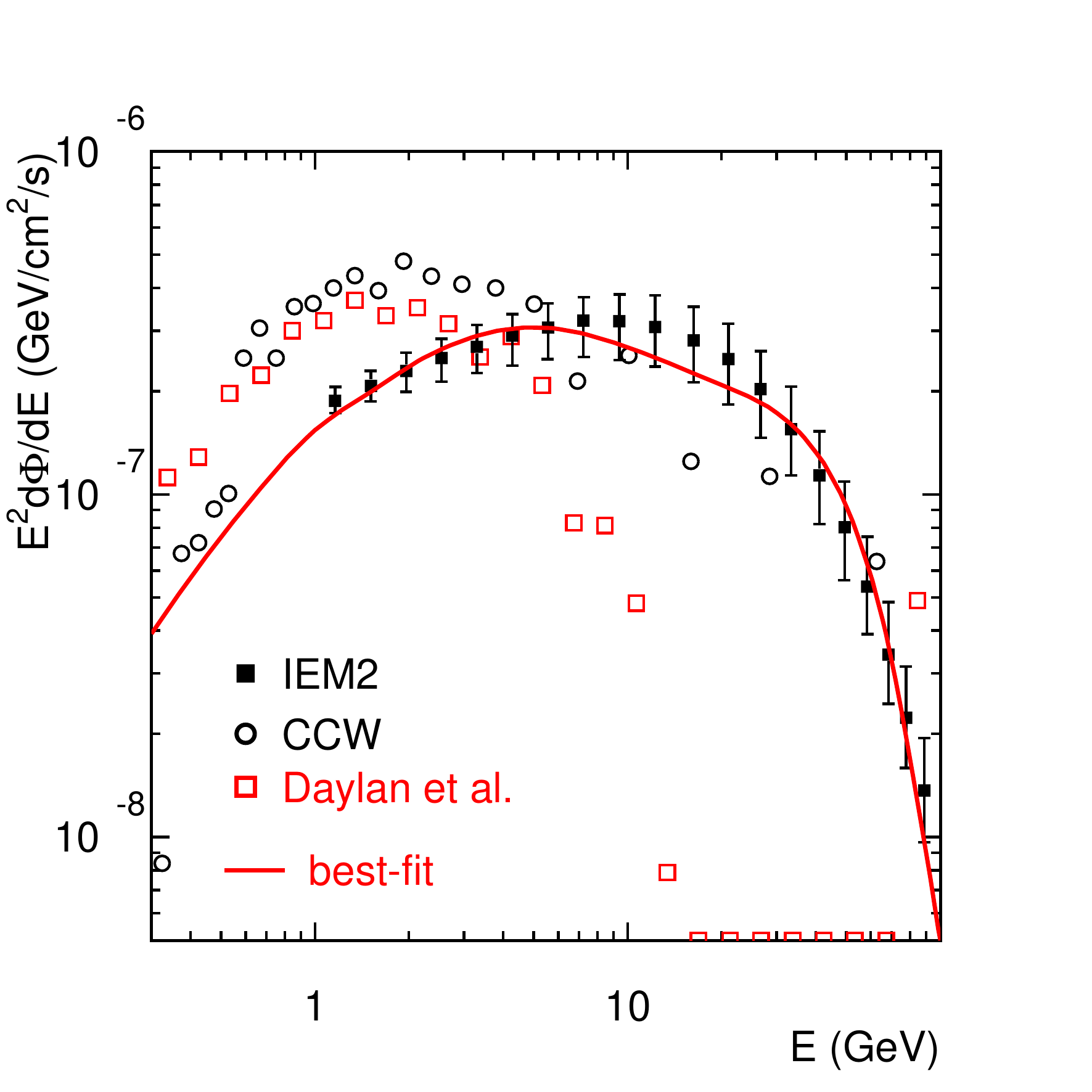}
\includegraphics[scale=1,width=5cm]{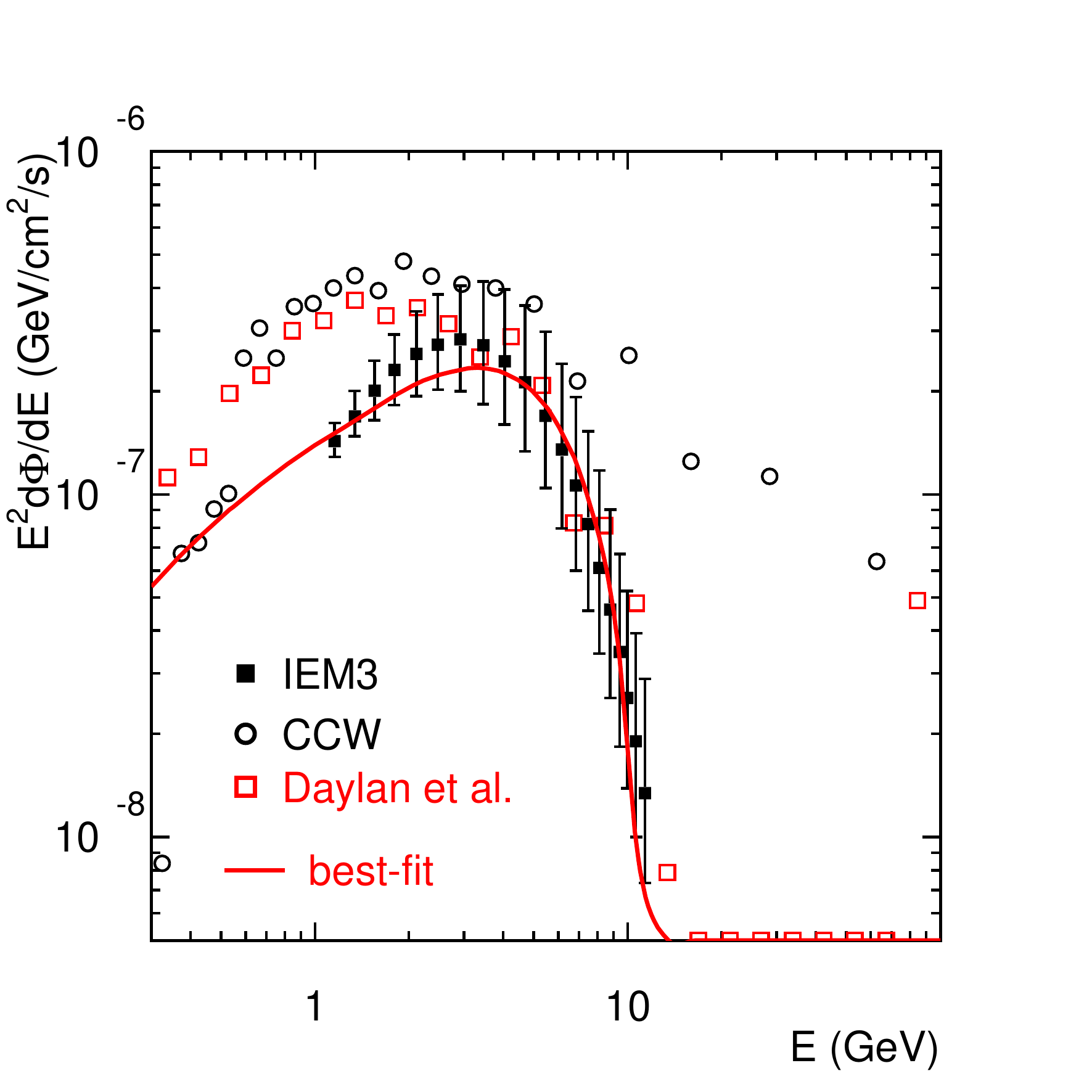}\\
\includegraphics[scale=1,width=5cm]{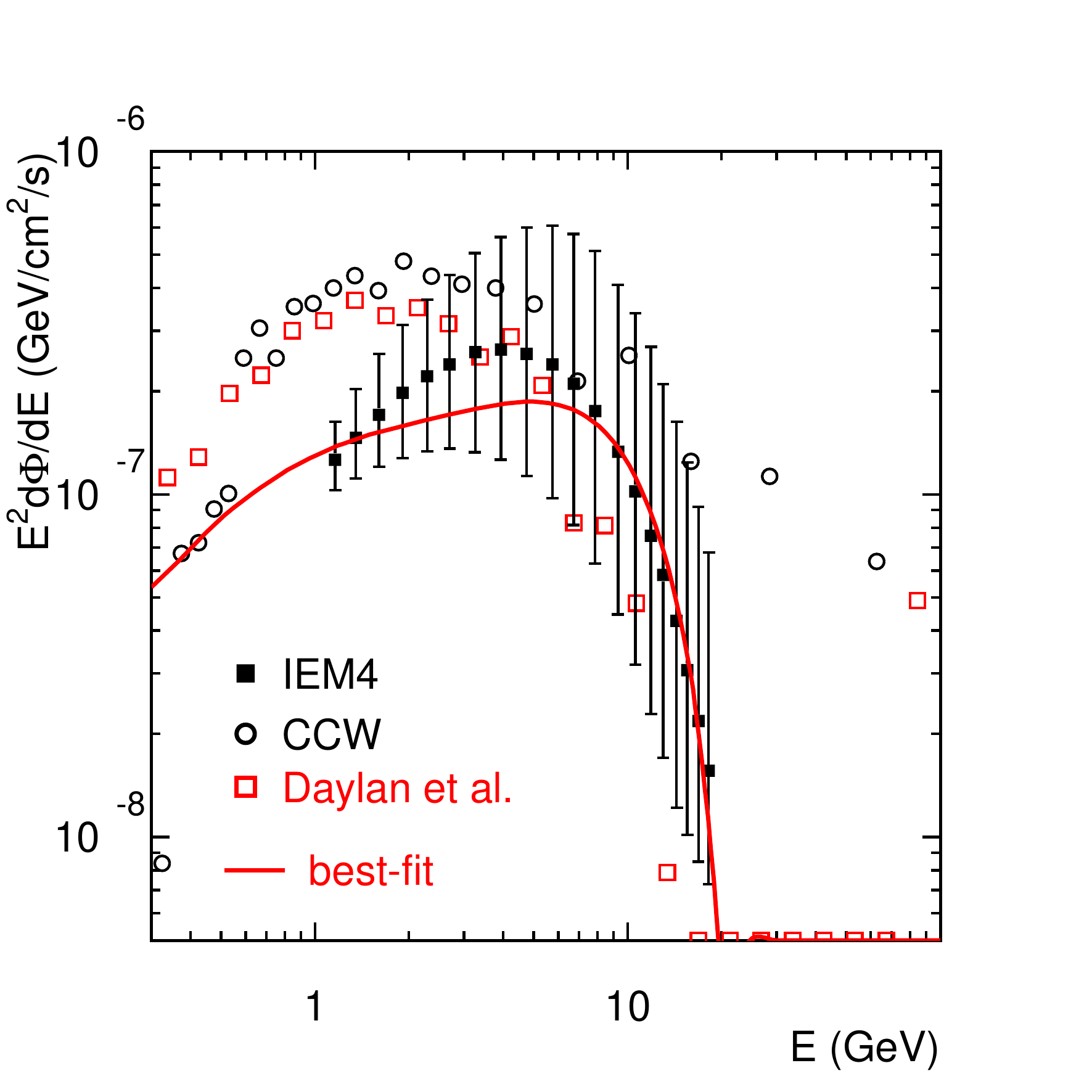}
\includegraphics[scale=1,width=5cm]{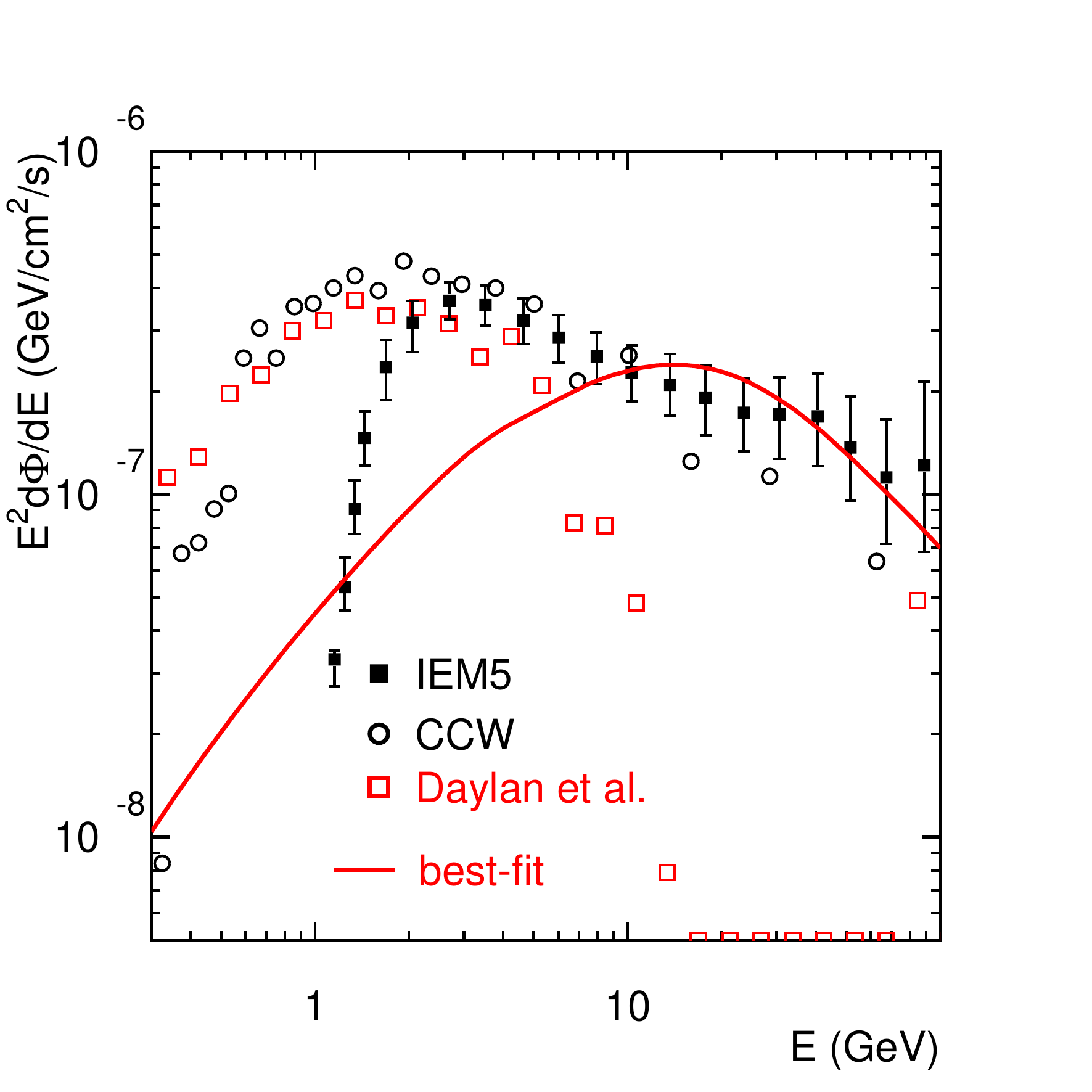}
\includegraphics[scale=1,width=5cm]{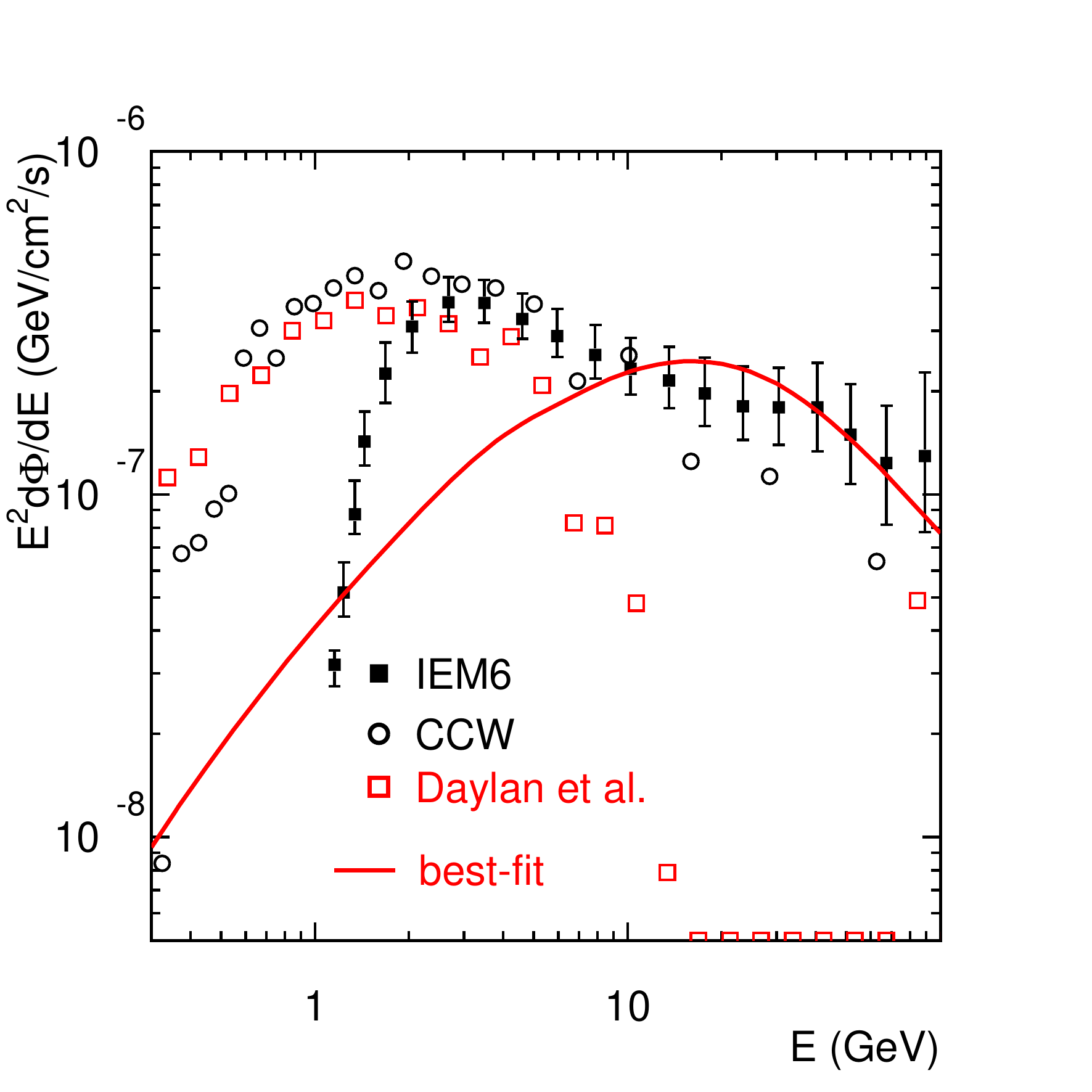}\\
\includegraphics[scale=1,width=5cm]{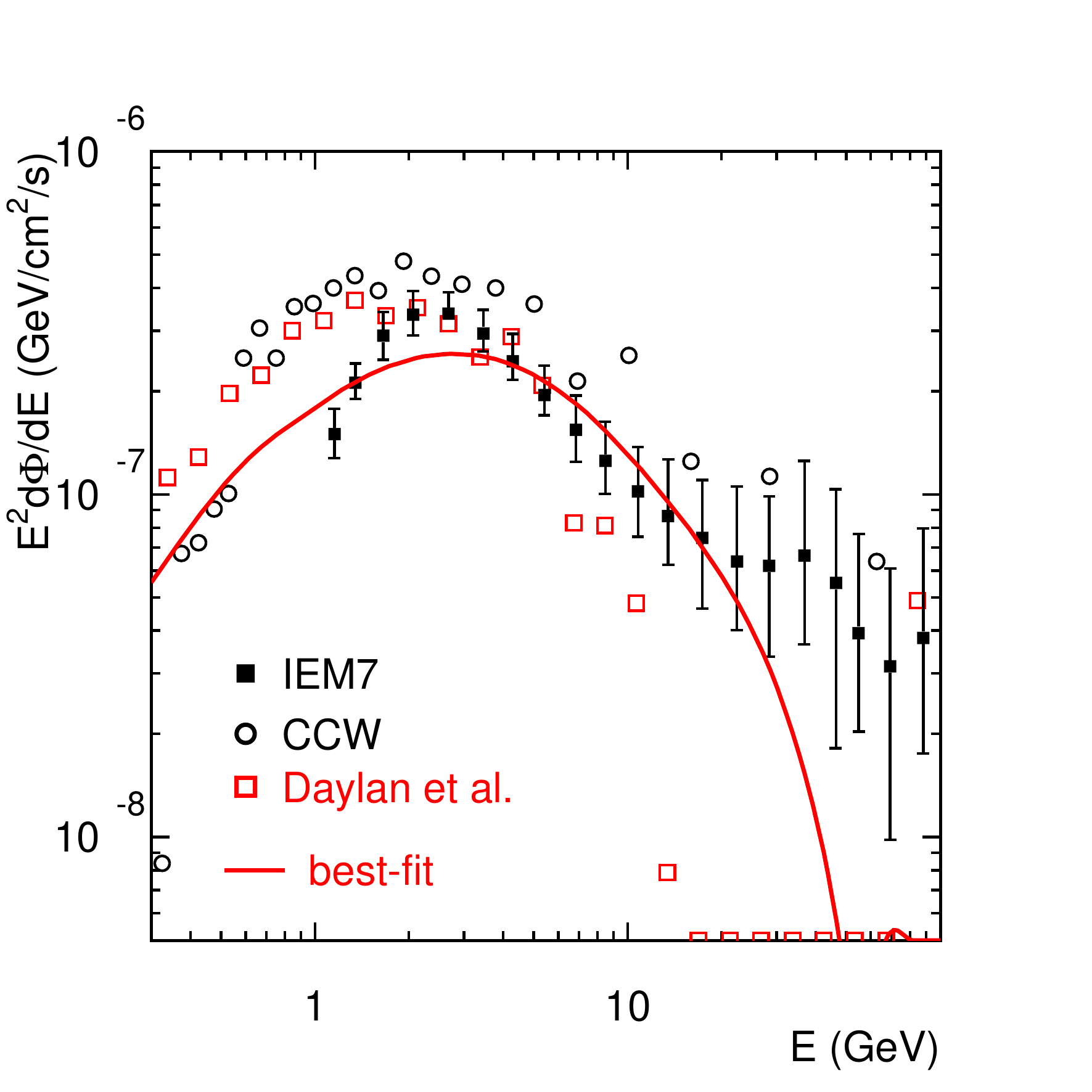}
\includegraphics[scale=1,width=5cm]{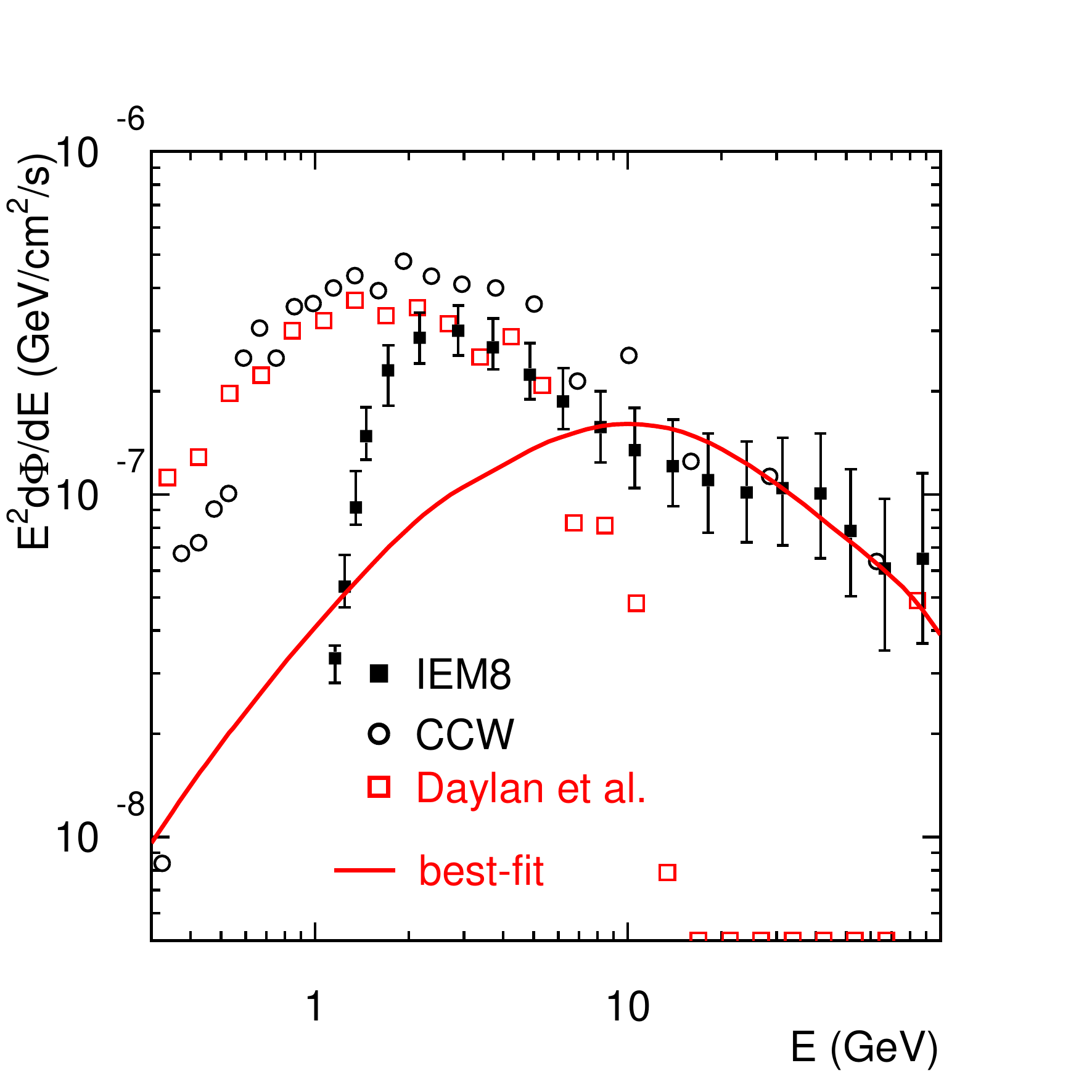}
\end{center}
\caption{Best fit results (red lines) of the simplified dark matter model to Fermi-LAT gamma ray residuals using the eight IEM scenarios.  The likelihood function includes the given Fermi-LAT IEM scenario (filled data points with error bars) and limits from spheroidal dwarf galaxies.  Data points from Daylan et al.~\cite{Daylan:2014rsa} and Calore et al. (CCW)~\cite{Calore:2014xka} are rescaled and shown as reference.}
\label{fig:gammarayfit}
\end{figure}

\begin{figure}[t]
\begin{center}
\includegraphics[scale=1,width=6.cm]{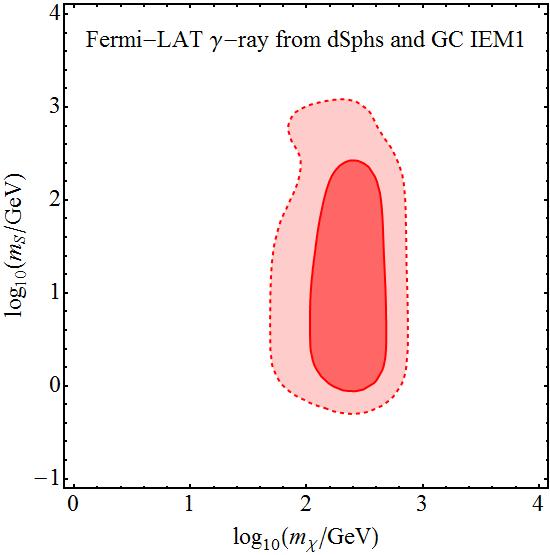}
\includegraphics[scale=1,width=6.cm]{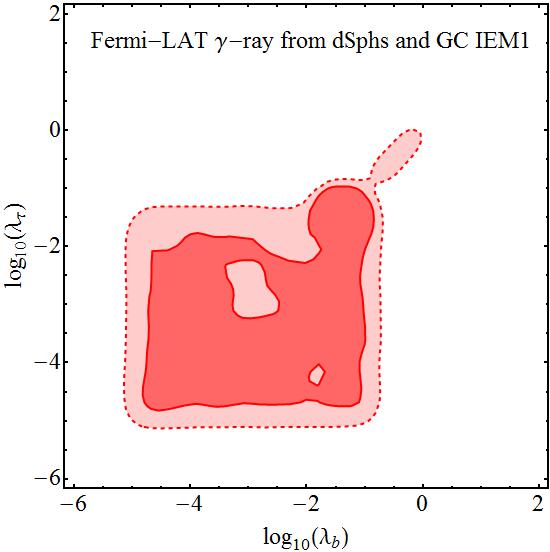}\\
\includegraphics[scale=1,width=6.cm]{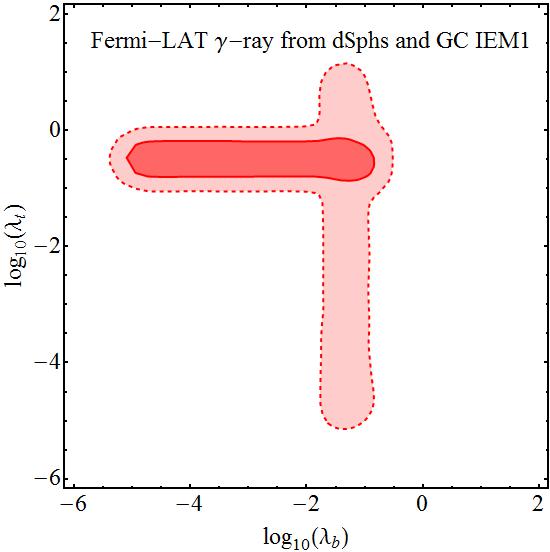}
\includegraphics[scale=1,width=6.cm]{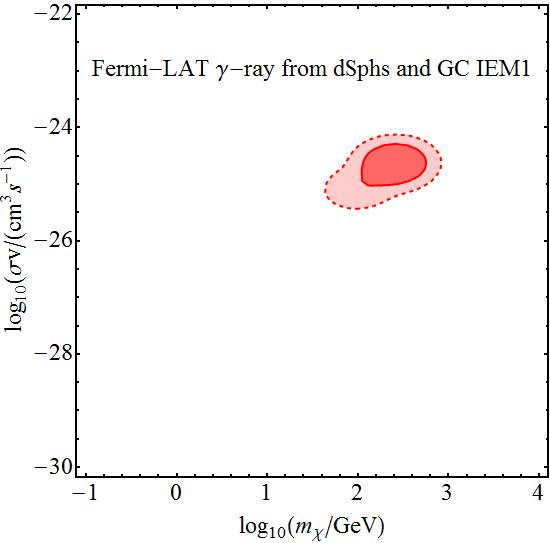}
\end{center}
\caption{Posterior probability distributions marginalized to the scanned model parameters. The likelihood function for these plots contains the Fermi-LAT gamma ray data from dwarf galaxies and Galactic Center IEM1. The dark and light regions hereinafter correspond to 68\% and 95\% credible regions, respectively.}
\label{fig:gammaraygcdwarf1}
\end{figure}

\begin{figure}[t]
\begin{center}
\includegraphics[scale=1,width=6.cm]{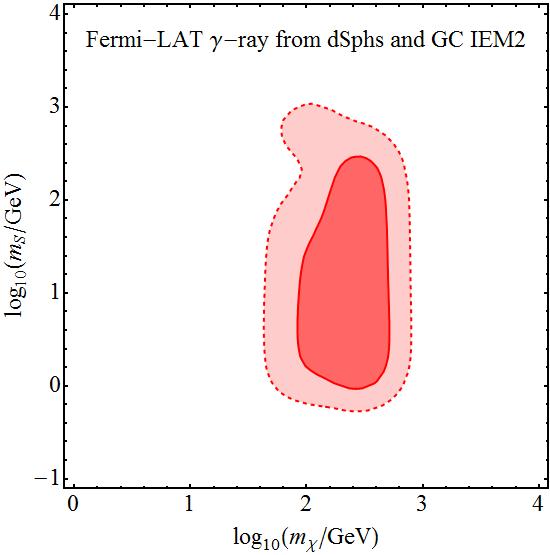}
\includegraphics[scale=1,width=6.cm]{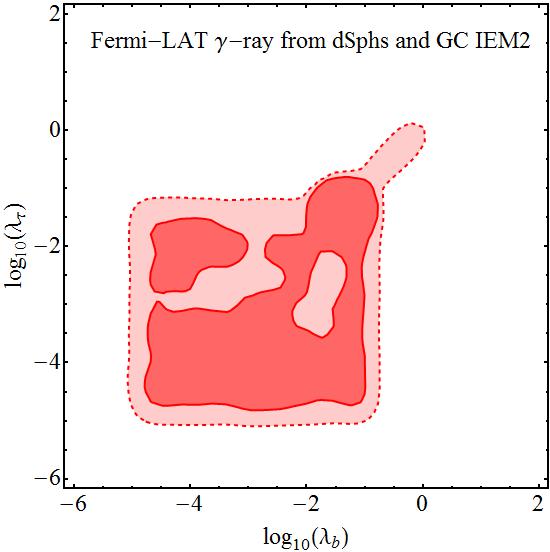}\\
\includegraphics[scale=1,width=6.cm]{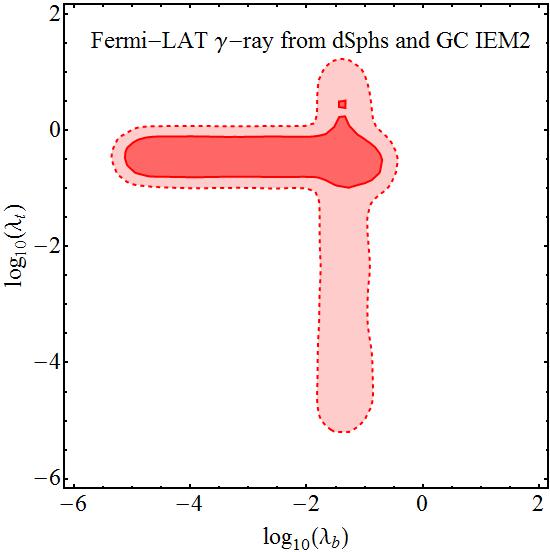}
\includegraphics[scale=1,width=6.cm]{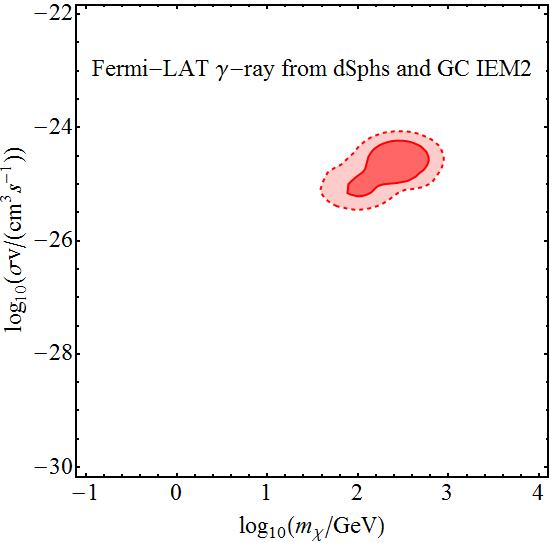}
\end{center}
\caption{Posterior probability distributions marginalized to the scanned model parameters. The likelihood function for these plots contains the Fermi-LAT gamma ray data from dwarf galaxies and Galactic Center IEM2.}
\label{fig:gammaraygcdwarf2}
\end{figure}

\begin{figure}[t]
\begin{center}
\includegraphics[scale=1,width=6.cm]{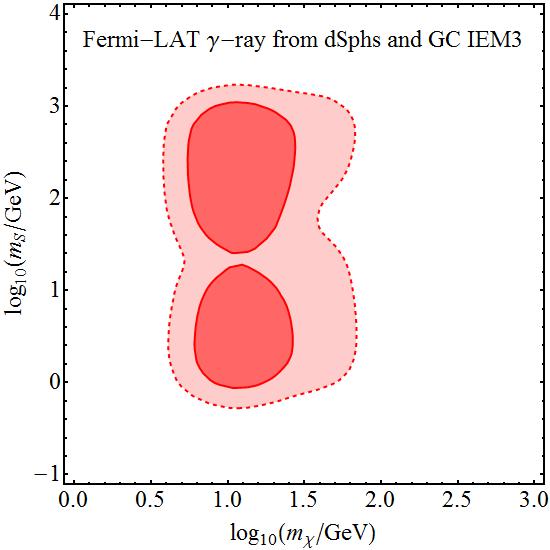}
\includegraphics[scale=1,width=6.cm]{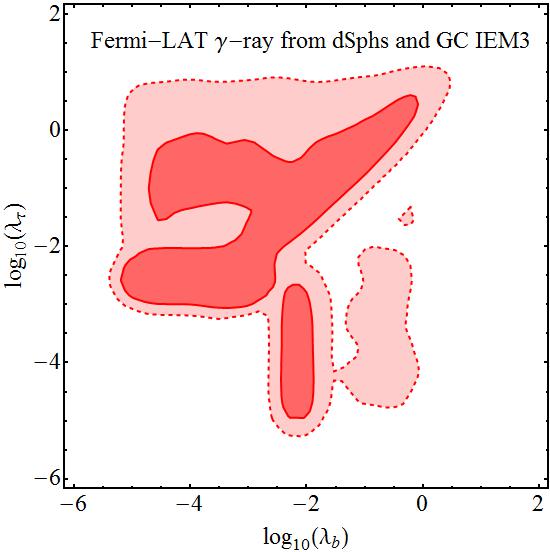}\\
\includegraphics[scale=1,width=6.cm]{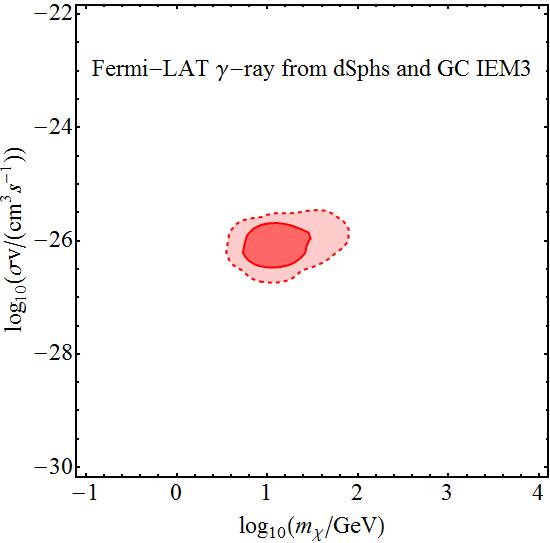}
\end{center}
\caption{Posterior probability distributions marginalized to the scanned model parameters. The likelihood function for these plots contains the Fermi-LAT gamma ray data from dwarf galaxies and Galactic Center IEM3.}
\label{fig:gammaraygcdwarf3}
\end{figure}

\begin{figure}[t]
\begin{center}
\includegraphics[scale=1,width=6.cm]{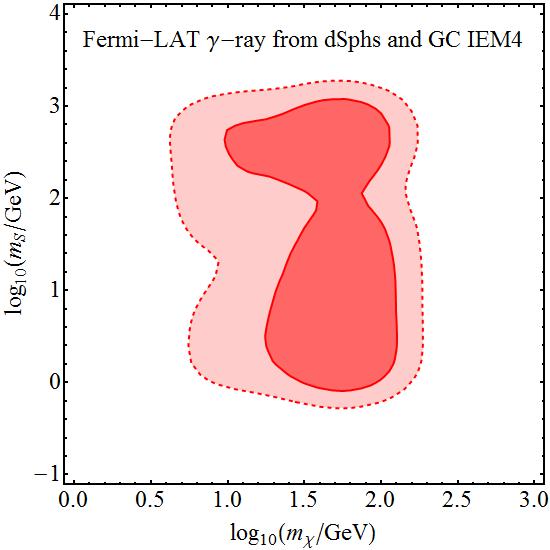}
\includegraphics[scale=1,width=6.cm]{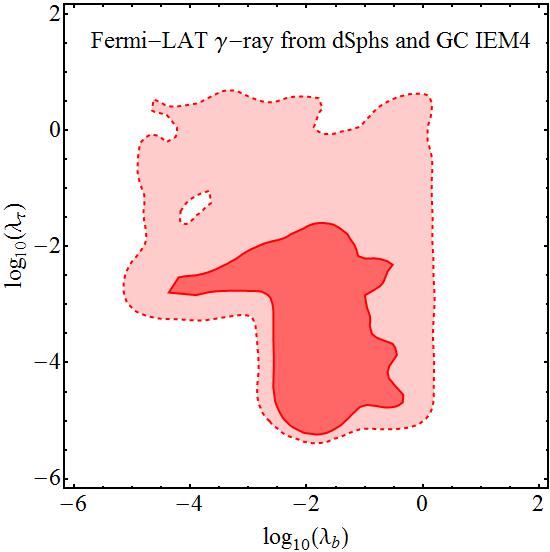}\\
\includegraphics[scale=1,width=6.cm]{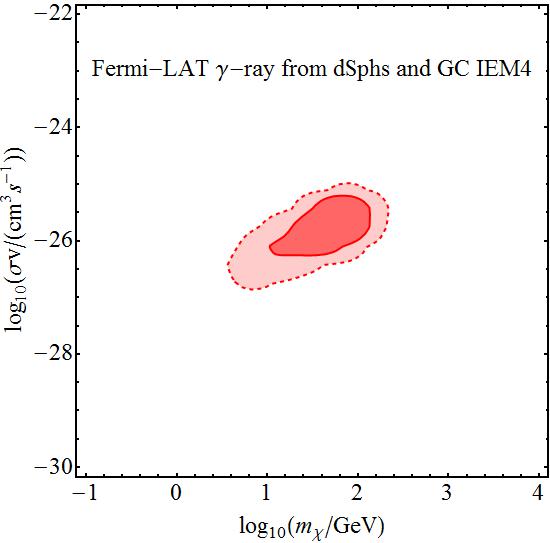}
\end{center}
\caption{Posterior probability distributions marginalized to the scanned model parameters. The likelihood function for these plots contains the Fermi-LAT gamma ray data from dwarf galaxies and Galactic Center IEM4.}
\label{fig:gammaraygcdwarf4}
\end{figure}

\begin{figure}[t]
\begin{center}
\includegraphics[scale=1,width=6.cm]{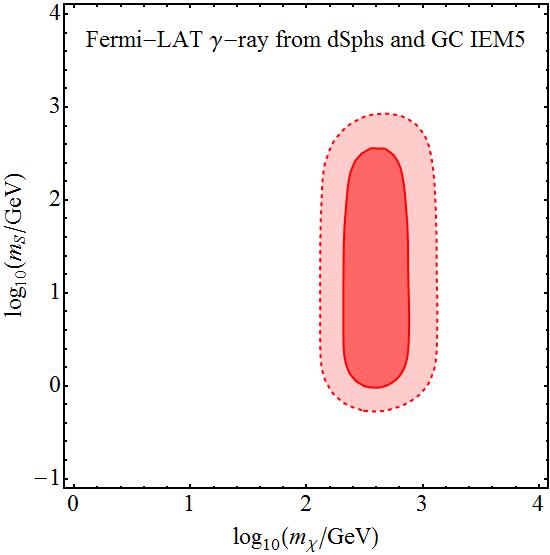}
\includegraphics[scale=1,width=6.cm]{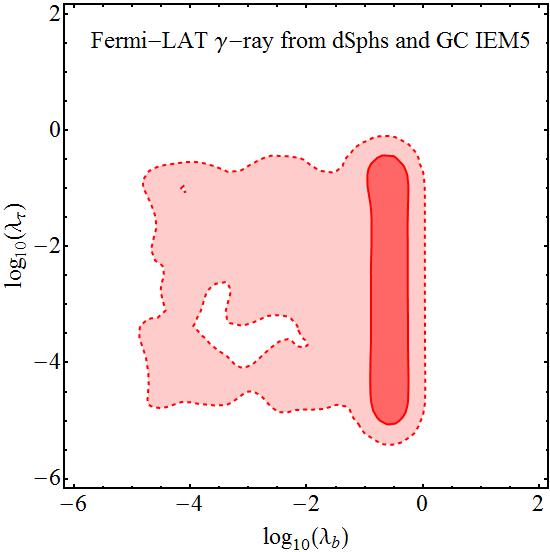}\\
\includegraphics[scale=1,width=6.cm]{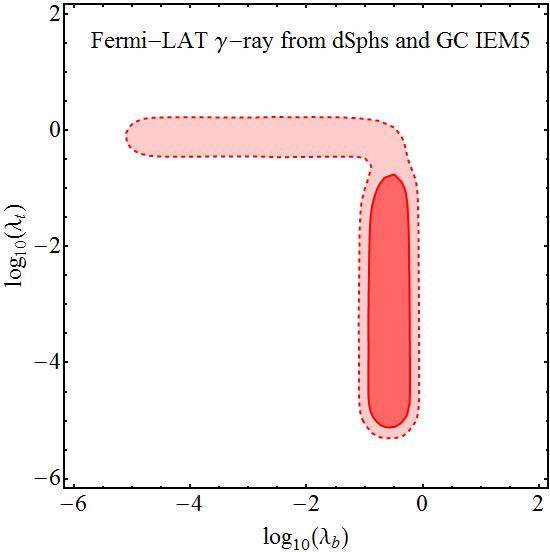}
\includegraphics[scale=1,width=6.cm]{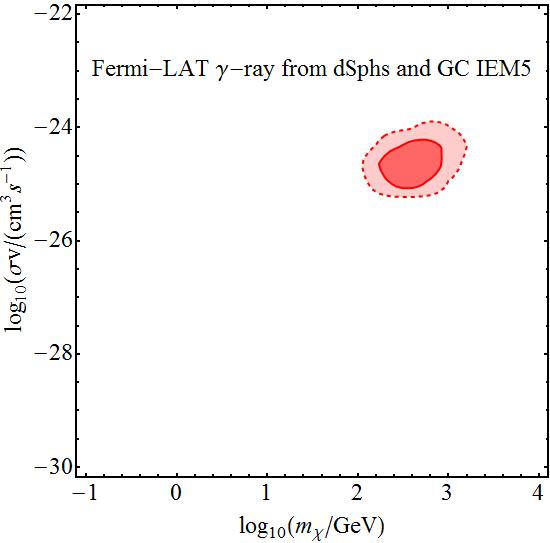}
\end{center}
\caption{Posterior probability distributions marginalized to the scanned model parameters. The likelihood function for these plots contains the Fermi-LAT gamma ray data from dwarf galaxies and Galactic Center IEM5.}
\label{fig:gammaraygcdwarf5}
\end{figure}

\begin{figure}[t]
\begin{center}
\includegraphics[scale=1,width=6.cm]{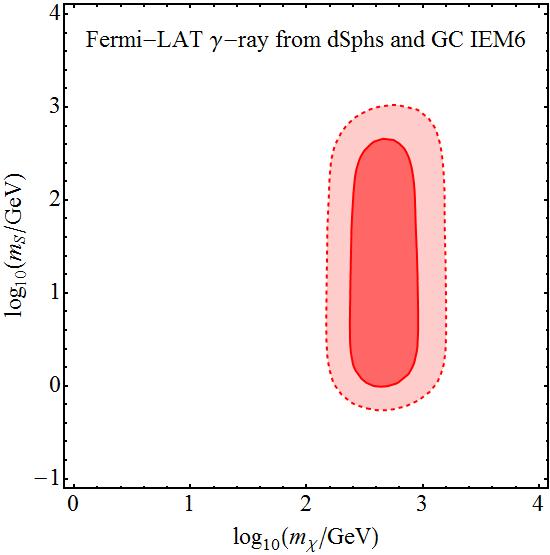}
\includegraphics[scale=1,width=6.cm]{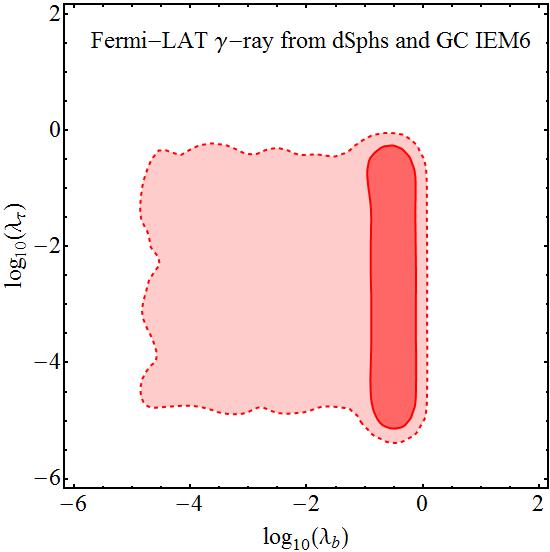}\\
\includegraphics[scale=1,width=6.cm]{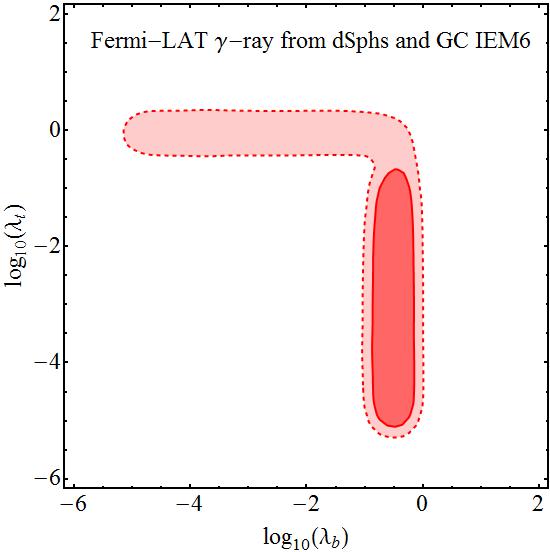}
\includegraphics[scale=1,width=6.cm]{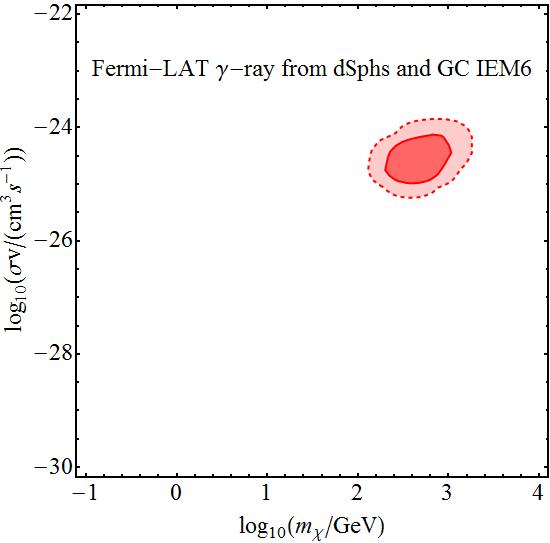}
\end{center}
\caption{Posterior probability distributions marginalized to the scanned model parameters. The likelihood function for these plots contains the Fermi-LAT gamma ray data from dwarf galaxies and Galactic Center IEM6.}
\label{fig:gammaraygcdwarf6}
\end{figure}

\begin{figure}[t]
\begin{center}
\includegraphics[scale=1,width=6.cm]{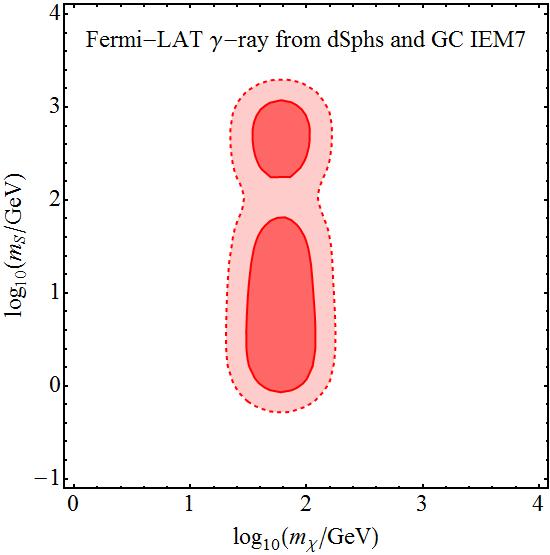}
\includegraphics[scale=1,width=6.cm]{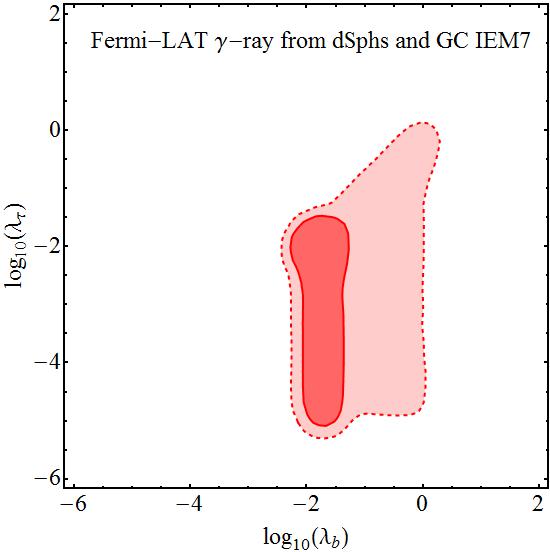}\\
\includegraphics[scale=1,width=6.cm]{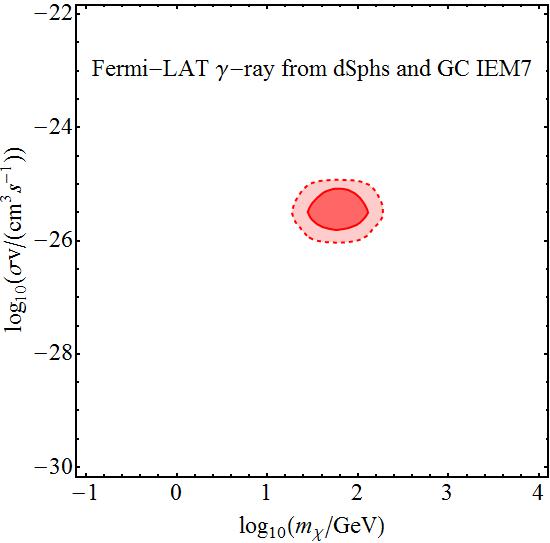}
\end{center}
\caption{Posterior probability distributions marginalized to the scanned model parameters. The likelihood function for these plots contains the Fermi-LAT gamma ray data from dwarf galaxies and Galactic Center IEM7.}
\label{fig:gammaraygcdwarf7}
\end{figure}

\begin{figure}[t]
\begin{center}
\includegraphics[scale=1,width=6.cm]{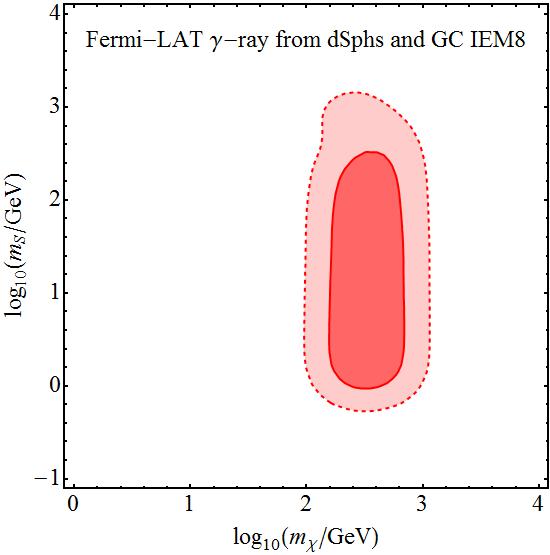}
\includegraphics[scale=1,width=6.cm]{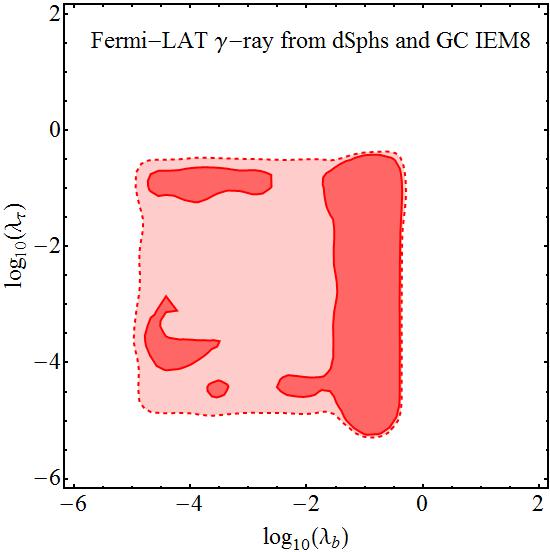}\\
\includegraphics[scale=1,width=6.cm]{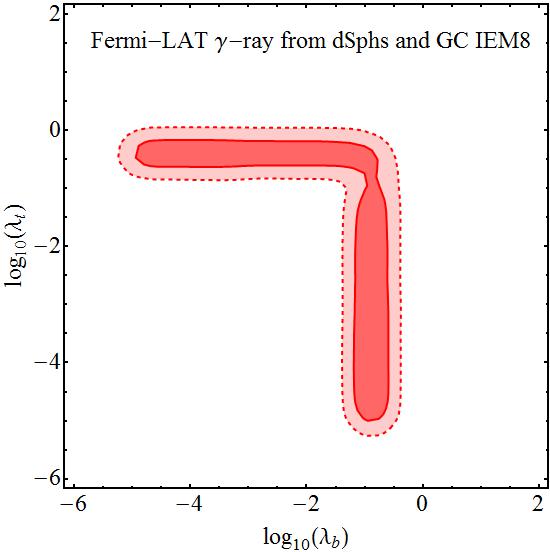}
\includegraphics[scale=1,width=6.cm]{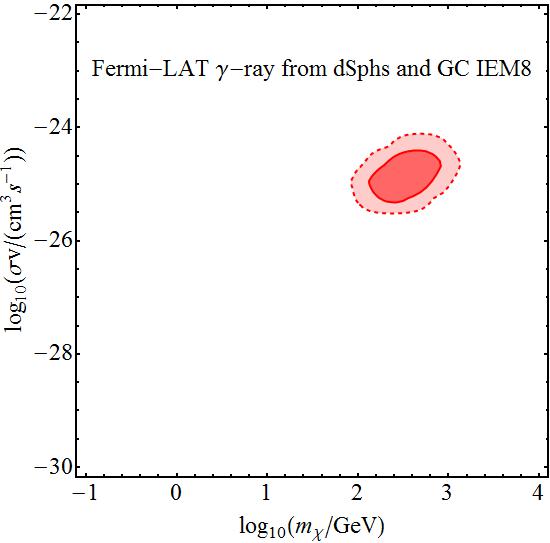}
\end{center}
\caption{Posterior probability distributions marginalized to the scanned model parameters. The likelihood function for these plots contains the Fermi-LAT gamma ray data from dwarf galaxies and Galactic Center IEM8.}
\label{fig:gammaraygcdwarf8}
\end{figure}

Table~\ref{bayes} shows that IEM scenarios 1, 2, 3 and 7 are the most compatible with the simplified dark matter model, while the rest (IEM4, 5, 6 and 8) relatively disfavor it.  Among the more congruous, the IEM2 scenario is slightly preferred over IEM1 which is somewhat preferred over IEM3.  IEM3 is moderately preferred over scenario 7.  These Bayes factors are in qualitative agreement with the best fits we find for the eight scenarios.  The latter are shown in FIG. 1.
As a byproduct of extracting the evidences for the eight Fermi IEM scenarios, we obtain posterior probability distributions of the dark matter theory parameters.  We marginalized these distributions to parameter pairs and plotted them in FIGs. \ref{fig:gammaraygcdwarf1}-\ref{fig:gammaraygcdwarf8}.  The posterior distributions show that in scenarios IEM1 and IEM2, the most favored ones by the Bayes factors, dark matter to standard model couplings are dominated by the coupling to the top quark (FIGs.~\ref{fig:gammaraygcdwarf1} and \ref{fig:gammaraygcdwarf2}).  Scenarios IEM3 and IEM4 both favor mixed bottom quark and tau lepton couplings (FIGs.~\ref{fig:gammaraygcdwarf3} and \ref{fig:gammaraygcdwarf4}). IEM4, however, is relatively disfavored partly due to the corresponding large uncertainties in the extracted Fermi distribution as seen in the fourth frame of FIG.~\ref{fig:gammarayfit}.  Scenario IEM7 favors pure $b$ quark coupling (FIG.~\ref{fig:gammaraygcdwarf7}) and coincides with the dark matter annihilation model singled out by Daylan et al.~\cite{Daylan:2014rsa} as the explanation of the gamma ray excess.  In the rest of the scenarios, (IEM5, IEM6, IEM8) which possess relatively low evidence, dark matter annihilation is governed by either $b$ or $t$ quark final states (FIGs.~\ref{fig:gammaraygcdwarf5}, \ref{fig:gammaraygcdwarf6} and \ref{fig:gammaraygcdwarf8}).  More complicated dark matter models are needed to fit these scenarios better.

Based on the Bayes factors and posterior distributions we can summarize our findings by grouping the the eight Fermi IEM scenarios as follows.
%
%
\begin{itemize}
\item IEM1, IEM2: the most preferred (intensity-scaled pulsars or OB stars IEM with exponentially cut-off NFW).  Dark matter annihilation is dominated by $t\bar{t}$ final state and the preferred DM mass is $\mathcal{O}(100)$ GeV.
\item IEM3: slightly less preferred (index-scaled pulsars IEM with exponentially cut-off NFW profile).  Dark matter annihilation final state is mixed $b\bar{b}$ and $\tau\bar{\tau}$ and the preferred DM mass is $\mathcal{O}(10)$ GeV.
\item IEM7: modestly preferred (index-scaled pulsars IEM with power law per energy band NFW profile).  Dark matter annihilation final state is dominantly $b\bar{b}$ with a preferred DM mass in the $\mathcal{O}(10)-\mathcal{O}(100)$ GeV region.  (The model of Daylan et al.~\cite{Daylan:2014rsa}.)
\item IEM4, IEM5, IEM6, IEM8: disfavored by our selected simplified dark matter model (which is most favored by other astrophysical data).
\end{itemize}

\section{Conclusions}
\label{sec:Concl}

In this work we inferred the relative plausibility of the eight interstellar emission models (IEMs) put forward by Fermi-LAT to interpret the gamma ray anomaly from the region of the Galactic Center.  Together with the most restrictive dwarf sheriodal observations, we included the Fermi-LAT data extracted from the eight IEM scenarios in a likelihood function.  As a dark matter hypothesis we selected the simplified model which emerged as the most plausible in previous studies.  This particle physics model played the role of the theoretical prediction in our likelihood function.  Using simple priors in the dark matter parameter space we calculated evidences for the eight IEMs and formed Bayes factors from them.  Comparing Bayes factors we found four of the IEMs compatible with the simplified dark matter model.  These four IEMs, however, imply fairly different mass range and annihilation final state for the dark matter particle.  Thus, unless the interstellar emission model is pinned down more precisely, the properties of dark matter extracted from gamma ray data will suffer from large uncertainties.

\acknowledgments
This work in part was supported by the ARC Centre of Excellence for Particle Physics at the Terascale.
The use of the Monash Advanced Research Computing Hybrid (MonARCH), the Multi-modal Australian ScienceS Imaging and Visualisation Environment (www.MASSIVE.org.au), and the National Computational Infrastructure (NCI), the Southern Hemisphere's fastest supercomputer, is also gratefully acknowledged.

\appendix

\section{Bayesian inference}

This Appendix provides a summary of the statistical background we rely on.  Possessing some prior information $I$, we denote the plausibility of two non-exclusive propositions $A$ and $B$ by $P(A|I)$ and $P(B|I)$.
In the context of our paper $A$ represents the gamma ray flux predicted by the dark matter model and $B$ the gamma ray residuals implied by the eight different IEMs.
The conditional expression
\begin{equation}
 P(AB|I) = P(A|BI) P(B|I) ,
\end{equation}
gives the probability that both $A$ and $B$ are correct.
From the symmetry of the above condition under the exchange of $A$ and $B$ follows Bayes theorem:
\begin{equation}
 P(A|BI) P(B|I) = P(B|AI) P(A|I) .
\end{equation}
We refer to $P(A|I)$ as the prior probability, the plausibility of proposition $A$ given information $I$, prior to $B$.  The likelihood function $P(B|AI)$ quantifies the plausibility of the occurrence of $B$ given $A$ and $I$.  The posterior $P(A|BI)$ indicates the probability of hypothesis $A$ given the data $B$.  The evidence $P(B|I)$ normalizes the posterior such that the latter can be interpreted as a probability.

For parametric theories, continuously spanned over a parameter space ${\pazocal p} = \{{\pazocal p}_1,...,{\pazocal p}_n\}$, Bayes theorem is written in the form
\begin{equation}
 \mathcal{P}({\pazocal p}|BI) {\cal E}(B|I) = \mathcal{L}(B|{\pazocal p}I)\Pi({\pazocal p}|I) ,
\label{eqnBayes}
\end{equation}
with the likelihood fucntion $\mathcal{L}(B|{\pazocal p}I)$ and the prior $\Pi({\pazocal p}|I)$ becoming distributions over ${\pazocal p}$.
The evidence
\begin{equation}
 {\cal E}(B|I) = \int_{D\pazocal p} \mathcal{L}(B|{\pazocal p}I)\Pi({\pazocal p}|I) \prod_{i=1}^{n} d{\pazocal p}_i ,
\end{equation}
is the integral of the posterior distribution over the full domain of ${\pazocal p}$ and specifies the marginalized probability of both propositions $A$ and $B$ being correct.
The marginalized posterior density
\begin{equation}
\mathcal{P}({\pazocal p}_1{\pazocal p}_2|BI) =
\int_{D\pazocal p} \mathcal{L}(B|{\pazocal p}I)\Pi({\pazocal p}|I) \prod_{i=3}^{n} d{\pazocal p}_i ,
\label{Eq:margin}
\end{equation}
is used to define credibility regions that include 68 or 95 percent of the probability mass.

The likelihood function in Eq.~(\ref{eq:LGalaCent}) is defined as
\begin{equation}
\mathcal{L}^{\rm Gauss} \left(\left. d_{i_{\rm bin}} \right| \pazocal p \right) = \frac{1}{\sqrt{2\pi}\sigma}\mathrm{Exp}\left(-\frac{(d_{i_{\rm bin}} - t({\pazocal p}))^2}{2\sigma^2}\right) ,
\end{equation}
where $d_{i_{\rm bin}}$ is the experimental measurement in the $i^{th}$ bin and $t({\pazocal p})$ is the theoretical prediction corresponding to the middle of the same bin.  The width of the Gaussian is given by the experimental and theoretical uncertainties combined in quadrature.
For Eq.~(\ref{eq:Ldwarf}) the likelihood function is a complementary error function
\begin{equation}
\mathcal{L}_i^{\rm Err} \left(\left. d_{i_{\rm bin}} \right| \pazocal p \right) =
\frac{1}{2}\mathrm{Erfc} \left(- \frac{d_{i_{\rm bin}} - t({\pazocal p})}{\sigma}\right) ,
\end{equation}
with the above defined $d_{i_{\rm bin}}$, $t({\pazocal p})$ and $\sigma$.


\end{document}